\documentclass[12pt]{article}
\usepackage{amsfonts,amsmath,amssymb,indentfirst}
\usepackage{graphicx,color}
\numberwithin{equation}{section}
\newcommand{\de}{\partial}
\newcommand{\be}{\begin{equation}}
\newcommand{\ba}{\begin{eqnarray}}
\newcommand{\ea}{\end{eqnarray}}
\newcommand{\ee}{\end{equation}}

\newcommand{\f}{\frac}
\newcommand{\s}{\sqrt}
\newcommand{\vp}{\varphi}

\newcommand{\ti}{\tilde}
\newcommand{\ap}{\alpha}

\newcommand{\ddd}{\cdot\cdot\cdot}
\newcommand{\no}{\nonumber \\}
\newcommand{\la}{\langle}
\newcommand{\lb}{\rangle}

\newcommand{\ov}{\overline}

\begin{document}

\begin{titlepage}
\thispagestyle{empty}
\renewcommand{\thefootnote}{\fnsymbol{footnote}}

KUNS-2160

KEK-TH-1281

IPMU-08-0080


\begin{center}
\noindent{\large \textbf{Disordered Systems and
the Replica Method in AdS/CFT}}\\
\vspace{1.5cm} \noindent{

Mitsutoshi Fujita$^{a}$\footnote{e-mail: mfujita@gauge.scphys.kyoto-u.ac.jp},\!
Yasuaki Hikida$^{b}$\footnote{e-mail: hikida@post.kek.jp},\!
Shinsei Ryu$^{c}$\footnote{e-mail: sryu@berkeley.edu}},\!
Tadashi Takayanagi$^{d}$\footnote{e-mail: tadashi.takayanagi@ipmu.jp}\\
\vspace{1cm}

 {\it $^{a}$Department of Physics, Kyoto University, Kyoto 606-8502, Japan \\
 $^{b}$High Energy Accelerator Research Organization (KEK), \\
 Tsukuba, Ibaraki 305-0801, Japan\\
 $^{c}$Department of Physics, University of California, Berkeley, CA 94720, USA\\
 $^{d}$Institute for the Physics and Mathematics of the Universe (IPMU), \\
 University of Tokyo, Kashiwa, Chiba 277-8582, Japan}
\end{center}
\vskip 2em

\begin{abstract}

We formulate a holographic description of
effects of disorder in conformal field theories
based on the replica method and the AdS/CFT correspondence.
Starting with $n$ copies of conformal field theories,
randomness with a gaussian distribution is described by a deformation
of double trace operators.
After computing physical quantities, we take the $n\to 0$ limit at the
final step.
We compute correlation functions in the disordered systems
by using the holographic replica method as well as the formulation in
the conformal field theory.
We find examples where disorder changes drastically
the scaling of two point functions.
The renormalization group flow of
the effective central charge in our disordered systems is also discussed.

\end{abstract}

\setcounter{footnote}{0}

\renewcommand{\thefootnote}{\arabic{footnote}}

\end{titlepage}

\newpage

\section{Introduction}
\label{introduction}

Classical statistical or quantum many-body systems can be studied
in the presence of
spatial inhomogeneity.
Indeed, it is quite rare that a perfectly clean system is realized in experiments,
and hence having a good understanding of effects of randomness/impurities is to some extent necessary.
Besides such a practical motivation,
disorder by itself or combined effects of disorder and interactions
can give rise to rich phenomena,
which deserve studies in their own right.
To name a few,
complex behaviors in spin glass systems
such as the Ising model
with random bonds (random ferromagnetic interactions)
or random magnetic field
\cite{Binder_Young_RMP},
or
Anderson localization of electronic systems
in the absence/presence of
electron-electron interactions
\cite{LeeRamakrishnan,BelitzKirkpatrick}
have been discussed.

When the amount of disorder is small or
disorder is (marginally) irrelevant in the
renormalization group (RG) sense,
effects of disorder can be studied perturbatively around a clean critical
point.
On the other hand, randomness is not necessarily small and it can drive
the system to a new type of disorder-dominated critical point,
called random critical point.
Examples include
a multicritical point in the random bond Ising model
\cite{Binder_Young_RMP,random_bond_Ising},
the integer quantum Hall plateau transition
\cite{Pruisken},
and a possibility of metal-insulator transitions
in (2+1)-dimensional correlated electron systems
\cite{Kravchenko}.
These putative critical points are beyond perturbative or mean-field treatments
and understanding the nature of these random critical points
has remained to this date as a major challenge in condensed matter physics.

It is the purpose of this paper to discuss quenched disordered systems
in the framework of AdS/CFT correspondence \cite{Ma}.
Effects of disorder have been studied as a perturbation
to weakly coupled and exactly solvable (conformal) field theories,
in terms of the perturbation theory or perturbative RG,
assuming the disorder strength is small
(see, e.g., \cite{CL,Be,CardyBook,CardyR,GL}).
On the other hand,
if we apply the AdS/CFT correspondence, it might be
possible to solve strongly disordered problems because AdS/CFT
is a strong/weak duality.
In AdS/CFT setups,
CFTs are typically non-abelian gauge theories \cite{Ma} or the critical $O(N)$ vector model
\cite{KP}.
Even though their precise relations to real condensed matter systems are
not clear at present, recently
there have been several hopes and circumstantial evidences that the AdS/CFT
can capture essential features of condensed matter systems,
such as the electrical and thermal transport
\cite{CON},
the quantum Hall plateau transition \cite{Kr},
the superconductivity \cite{HTC}, the entanglement
entropy \cite{RT}, the scale invariant theories with non-trivial dynamical
exponents \cite{NRE},
and so on.
In the presence of weak randomness a holographic analysis for
disordered systems was given in \cite{HH}.

Assuming that disorder configurations such as
the location of impurities
are distributed according to some underlying probability distribution,
quantities of our interest (e.g., local correlation functions)
also fluctuate from different disorder configurations.
We are thus forced to deal with
the probability distribution of observables
or in particular the first (or first few) moment(s) of the observables.
At this point, it is important to emphasize that
the average over disorder configurations is taken \textit{after}
we take the statistical mechanical ensemble average
over spins,
or the path integral over (quantum) field configurations and so on.
This is a major difficulty in disordered systems
since for a given distribution of disorder
we do not have translation invariance,
although correlation functions after quenched disorder averaging
may be translation invariant.
A standard tool to analyze effects of disorder is \textit{the replica method
(or replica trick)}
(see, e.g., \cite{ill, Mezard, CardyBook} and section
\ref{sec: Disordered Systems and Replica Method} in the present paper).

In this paper, we consider a CFT deformed
by a certain operator with its coupling changing from position to position.
We will show that with the replica method,
a generalization of the double trace interaction \cite{ABS,Wi} can
nicely describe the randomness. Based on this idea,
we will formulate the holographic replica method with some examples
in section \ref{Disordered Systems via AdS/CFT}. In particular, we calculate the
two point functions and show that their scaling behaviors importantly change due to the
randomness.
In section \ref{Field Theory Analysis in the Planar Limit}, we will give a complementary field
theoretic analysis of the same
system and confirm that it agrees with the holographic result.
In appendix \ref{Order/Disorder Phase Transition}, we will present a generalized holographic replica model and
realize the order/disorder phase transition.

In contrast to clean systems,
the structure of the RG flows connecting random critical
points is less understood in disordered systems.
This is so since critical theories describing random critical
points are expected to be non-unitary.
This in turn means we cannot use, in two dimensions, say,
the $c$-theorem by Zamolodchikov,
to know the direction of the RG flow.
In particular, when the supersymmetric disorder averaging is applicable,
the central charge is always vanishing $c=0$ because of the
cancellation among matter fields and ghosts.
Gurarie \cite{Gurarie,GL} introduced,
with the use of the supersymmetric disorder averaging,
the effective central charge (or Gurarie's $b$)
and proposed to use it to measure the degrees of freedom of a disordered
two-dimensional CFT.
Later, the equivalent quantity
in the replica method (which we call $c_{eff}$ in this paper)
was introduced by Cardy \cite{Cardy}.
(See appendix \ref{apb}.)
It is an open problem if the effective central charge $c_{eff}$
shows the irreversible relation $c^{IR}_{eff}<c^{UV}_{eff}$ along the RG flow,
as the central charge in two-dimensional unitary CFTs does.
In appendix \ref{apb},
we take a first step of the holographic calculation
of $c_{eff}$ and show that the $c$-theorem like
relation $c^{IR}_{eff}<c^{UV}_{eff}$ holds in examples
where we can calculate $c_{eff}$ straightforwardly.

\section{Disordered Systems and the Replica Method}
\label{sec: Disordered Systems and Replica Method}

Let us start by reviewing
how the replica method can be applied
to quenched disordered systems
(see, for a review, e.g. \cite{ill, Mezard, CardyBook})%
\footnote{
There are two types of disorder:
annealed disorder and quenched disorder. The former is the case where impurities
are in thermal equilibrium with main system, while in the latter, it is not.
In this paper we only discuss the latter case. Refer to \cite{CardyBook} for more details.}.
Our arguments cover any
quantum field theory (QFT) and we simply represent the action of a $d$-dimensional
QFT as ${S}_{0}[\vp]=\int d^dx \, {\cal L}_{0}(\vp)$ with abstractly expressing all fields as
$\vp$.
We pick up a certain operator in this theory and denote it by
${\cal{O}}(x)$.
We perturb this
theory by adding an interaction of the form \be S={S}_{0}+\int
d^{d} x ~g(x){\cal{O}}(x). \label{lagone} \ee This defines a classical disordered system
in $d$ spacial dimensions.
In a $d+1$-dimensional quantum system,
a spatially inhomogeneous perturbation by disorder is given by \\
$\int dt d^dx \, g(x){\cal{O}}(x,t)$,
where $t$ represents the (imaginary) time direction, and
the disorder configuration $g(x)$ depends only on the spacial coordinates $x$.
Below we shall proceed in the
classical setup, though we can extend to the latter case straightforwardly.

A quenched disordered system (or random system) is such a system where the coupling $g(x)$
is depending randomly on the spatial coordinates $x$.
We assume that the randomness is distributed with a
gaussian profile, i.e., its distribution functional is given by
\be P[g(x)] \propto
e^{-\f{1}{2f}\int d^d x \, g(x)^2} \label{prob} \ee
with $f > 0$.
The free energy in the disordered system
$\ov{\log Z}$ can be found by simply taking the gaussian average of $\log Z$
over disorder,
where $\overline{\cdots}$ represents disorder averaging with respect
to the distribution $P[g(x)]$.

In disordered systems,
we are interested in
correlation functions averaged over disordered configurations.
An important point is that this random average is not equivalent to
taking the random average for the action itself (\ref{lagone}).
Instead the averaged correlation function is expressed as follows;
\begin{align}
&\overline{\la{\cal{O}}(x_1){\cal{O}}(x_2)\ddd {\cal{O}}(x_k)\lb}
 = \label{corone}
 \\ & \qquad = \int
Dg(x)P[g(x)]\left[\f{\int D\vp \, e^{-S_0[\vp] - \int
d^d x \, g(x){\cal{O}}(x)}{\cal{O}}(x_1){\cal{O}}(x_2) \ddd
{\cal{O}}(x_k)}{\int D\vp \, e^{-S_0[\vp] - \int
d^d x \, g(x){\cal{O}}(x)}} \right]. \nonumber
\end{align}
The most important technical problem is how to deal with the factor of the
inverse partition function. One way to overcome this difficulty is to
introduce ghosts and represent the inverse partition function by that of ghosts.
This is called the supersymmetric method
and at a random fixed point it is typically described by a non-unitary CFT
with the vanishing total central charge (see, e.g., \cite{Efetov,GL,Be}).
This procedure is quite useful since we can utilize the knowledge of
quantum field theory technique. However, it has a severe disadvantage
that presently it can be used only for limited situations, where the
original theory described by $S_0$ is a free field theory.

Instead of proceeding along this direction,
we would like to resort to another method called the replica method, which
can be applied to any quantum field theories
(refer to, for example, \cite{ill}).
The method may be summarized as follows.
First we introduce $n$ copies of the QFT and denote
the $i$-th copy as QFT$_i$.
Next we prepare $n$ copies of the field
$\vp_i$ in QFT$_i$. Then we consider the path integral in the
product of $n$ QFTs,
i.e. ${\mbox{QFT}}_1\otimes {\mbox{QFT}}_2\otimes \ddd \otimes {\mbox{QFT}}_n$, given by
\begin{align}
&\int Dg(x)P[g(x)]\prod_{i=1}^n [D\vp_i]~ e^{-\sum_{i=1}^n
S_0[\vp_i] - \int d^{d} x \, g(x)\sum_{i=1}^n
{\cal{O}}_i(x)}{\cal{O}}_1(x_1){\cal{O}}_1(x_2)\ddd
{\cal{O}}_1(x_k)\no & \qquad
={\int}
\prod_{i=1}^n [D\vp_i] \, e^{-\sum_{i=1}^n
S_0[\vp_i]+\f{f}{2}\int d^d x \left(\sum_{i=1}^n
{\cal{O}}_i(x)\right)^2}{\cal{O}}_1(x_1){\cal{O}}_1(x_2)\ddd
{\cal{O}}_1(x_k), \label{cortwoo}
\end{align}
where ${\cal{O}}_1(x)$ is the operator ${\cal{O}}(x)$ in the Hilbert space of
QFT$_1$.
In the replica method, we identify the average (\ref{corone}) with
the $n\to 0$ limit of (\ref{cortwoo}).
The parameter $n$ is originally a positive integer number,
but we assume the analytical continuity with respect to $n$
and take the limit $n\to 0$ finally.
Indeed the inverse partition function required in (\ref{corone}) formally
appears in this limit.

Let us define the scaling dimension of ${\cal{O}}(x)$ by $\Delta$,
then the deformation induced by the randomness is relevant or marginal if
\be
2\Delta \leq d . \label{relevant}
\ee
This condition is clear from
power counting in the replicated action (\ref{cortwoo}) and
is called Harris criterion.

We consider large $N$ gauge theories or a $O(N)$ vector model
 as examples of QFT since
they appear in AdS/CFT correspondence.
The operator ${\cal O}$ that couples to disorder $g(x)$
is then given by a single trace operator,
and with the replica trick the disorder effect is
expressed by the double trace deformations as in \eqref{cortwoo},
which have been studied in the context of AdS/CFT \cite{ABS,Wi,Mu}.
If we are interested in strongly coupled (or strongly
disordered)
regime of field theories,
then the path integral in \eqref{cortwoo}
would be quite difficult to compute.
Since the AdS/CFT correspondence maps strongly coupled CFTs to
weakly coupled gravity theories, it is desirable to establish
dual description to analyze strongly coupled disordered systems.
In this paper we mainly focus on the planar limit of large $N$
gauge theory, which is dual to
the classical limit of the dual gravity theory.

\section{Disordered Systems via AdS/CFT}
\label{Disordered Systems via AdS/CFT}

Now we move on to the main part of this paper: formulation of the AdS/CFT correspondence for
disordered systems. We consider a standard setup of AdS/CFT,
where a $d$ dimensional Euclidean CFT is dual to a gravity theory on
$d+1$ dimensional Euclidean anti-de Sitter (AdS) space
described by the metric
\be
ds^2=\f{dz^2+\sum_{\mu}dx^\mu dx_\mu}{z^2}. \label{adsmetric}\ee
The dual CFT is proposed to live on the boundary located at $z=0$.

A spin-less operator $\cal{O}$, which is typically a single trace operator in a gauge theory, is supposed to be dual to
a scalar field\footnote{We will stick to the scalar field example just for simplicity.
We can generalize this to, say, vector and spinor fields straightforwardly.} $\phi$ with a mass $m$
in the bulk AdS space. The relation between the conformal dimension $\Delta$ of $\cal{O}$ and the mass $m$ is given by the formula \cite{GKP,WAdS}
\be
\Delta_{\pm}=\f{d}{2}\pm\s{m^2+\f{d^2}{4}}. \label{dim}
\ee
For our purpose, we need a relevant operator in the replicated theory,
and hence we require \eqref{relevant}.
{}From this condition we pick up
the smaller dimension $\Delta_{-}$ in (\ref{dim}) and we call it simply $\Delta$ below.
The behavior of $\phi$ near the boundary $z=0$ looks like
\be
\phi(z,x)\sim z^{d-\Delta}(\alpha(x)+{
O}(z^2))+z^{\Delta}\left(\f{\beta(x)}{2\Delta-d}+{O}(z^2)\right). \label{expanda} \ee
In the standard interpretation of AdS/CFT \cite{KW},
$\ap(x)$ is regarded as a source to the dual operator ${\cal O}$,
while $\beta(x)$ is its expectation value
$\langle {\cal O} (x) \rangle = \beta (x)$.
We demand
the normalizability for the mode with
$\phi \sim z^{\Delta}$,
which leads to
a constraint on the range of $\Delta$.
Combining with \eqref{relevant} we have to choose%
\footnote{In order for $\Delta$ to satisfy this condition,
the mass of $\phi$ should be in the range of $- d^2 /4 < m^2 < (1 -d^2)/4$.
In AdS space tachyonic modes are allowed due to the curvature, and
the lower bound is known as Breitenlohner-Freedman bound \cite{BF}.}
\be
\f{d-2}{2}\leq\Delta \leq\f{d}{2}.\label{dimb}
\ee
The lower bound
is known to be dual to the unitarity bound in the dual CFT \cite{KW}.

For the scalar field $\phi$ we require the regularity at $z=\infty$,
then $\ap(x)$ and $\beta(x)$ are related as follows
\be \beta(x)=\pi^{-d/2}
\f{(2\Delta-d)\Gamma(\Delta)}{\Gamma(\Delta-d/2)}\int d ^d x'
\f{\ap(x')}{|x-x'|^{2\Delta}} . \label{bbeta}\ee
In the momentum space representation it can be expressed as
\be \beta(k)=G(k)\ap(k),\label{abg}
\ee where $G(k)$ is given by\footnote{We employed the formula \be \int
d^d x \f{e^{ikx}}{x^{2\Delta}}=2^{d-2\Delta}\pi^\f{d}{2}
\f{\Gamma(d/2-\Delta)}{\Gamma(\Delta)}k^{2\Delta-d}.\ee}
\be
G(k)=\f{(2\Delta-d)\Gamma(d/2-\Delta)}{\Gamma(\Delta-d/2)}\left(\f{k}{2}\right)^{2\Delta-d}.
\ee
Since we chose $\Delta_-$ instead of $\Delta_+$ as the conformal
dimension of dual operator ${\cal O}$, the term proportional to
$\ap(x)$ is less singular than the one proportional to $\beta(x)$ in
the boundary limit as opposed to the usual cases.
Despite this fact we can still
treat the former as a source because $\ap(x)$ and $\beta(x)$ are related to each other via a canonical transformation
as first claimed by Klebanov and Witten \cite{KW}.

\subsection{Double Trace Deformation in AdS/CFT}
\label{Double Trace Deformation in AdS/CFT}

According to \cite{Wi} (see also \cite{Mu})
a multi-trace deformation
$\int d^d x \, W ({\cal O} ) $
can be incorporated into the formulation of AdS/CFT correspondence.
Here we focus on the deformation of a CFT by a double trace
operator of the form ($\lambda>0$)
\be
S_{int} =\f{\lambda}{2}\int d^d x \,
 ( {\cal O}(x) )^2 ,
\label{dti} \ee
as a preparation for the later analysis of disordered systems.
Assigning the boundary behavior \eqref{expanda} and demanding
the regularity at
$z \to 0$, the scalar field is uniquely parameterized by $\alpha (x)$
or $\beta (x)$. As mentioned before these two parameters can be
exchanged by a Legendre transform, and it is useful for our purpose
to use $\beta (x)$ instead of $\alpha (x)$.
Inserting the scalar field into the
kinetic term and partially integrating over the coordinate $z$,
we can obtain the action in terms of $\alpha (x)$. 
Then the Legendre transform leads to
\be
S[\beta]=\f{1}{2}\int d^d k \,
\f{\beta(k)\beta(-k)}{G(k)}. \ee
Notice that the result is expressed as a field theory on the boundary of AdS
space.

Recall that the expectation value of ${\cal O}(x)$ corresponds to
the variable $\beta (x)$.
Then the total action $S$ in the presence of the double trace deformation \eqref{dti} and a source to $\phi$
is expressed as \be S[\beta,J]=\int
d^d k \, \left[\f{1}{2}\beta(k)\left(\f{1}{G(k)}+\lambda \right)\beta(-k)+\beta(-k)J(k)\right].\label{saction}
\ee The equation of motion for $\beta$ leads to
\be
\f{1+\lambda
G(k)}{G(k)}\cdot\beta(k)+J(k)=0.
\label{eomj}
\ee
In other words, the boundary condition
for $\phi$ is now changed into $\ap+\lambda \beta+J=0$.
{}From \eqref{eomj} we can express
\be
S[J]=-\f{1}{2}\int d^d k \,
J(k)\left(\f{G(k)}{1+\lambda G(k)}\right) J(-k),
\ee
and in the end we obtain the two point function
\be \left.
\frac{\delta^2}{\delta J(k) \delta J(-k)} e^{-S [J]} \right|_{J=0}
= \la {\cal O}(k){\cal O}(-k) \lb
=\f{G(k)}{1+\lambda G(k)} \label{twopp}
\ee
taking the derivatives with respect to the source $J(k)$.

Due to the deformation of the (marginally) relevant operator \eqref{dti},
the conformal symmetry
should be broken in the CFT side.
Nevertheless, the vacuum solution in the AdS side
remains
trivial $\phi(x,z)=0$
even in the presence of non-vanishing $\lambda$. Hence the background
is still $AdS_{d+1}$, which implies the
conformal symmetry of dual CFT. This is so since
our calculation neglects backreaction in
the gravity theory, which should be taken into account from
one-loop order \cite{GM}.
Notice that the two point function (\ref{twopp})
breaks the conformal invariant%
\footnote{At the planar limit generic
correlation functions do not receive any contributions
from the deformation of double trace
operator. This fact can be understood by explicitly
writing down the double line Feynman diagrams.
The correlation functions involving ${\cal O}$ are exceptions.}.
In the UV limit $k\to \infty$, it is
approximated by \be \la {\cal O}(k){\cal O}(-k) \lb\sim G(k)\sim
k^{2\Delta-d}, \ee while in the IR limit we find \be \la {\cal
O}(k){\cal O}(-k) \lb\sim \f{1}{\lambda }-\f{1}{\lambda^2G(k)}\sim
\mbox{const.}+O(k^{d-2\Delta}) . \ee
This shows that under the RG flow
the operator with the conformal dimension $\Delta \, (=\Delta_-)$ flows to the one
with the conformal dimension $d-\Delta \, (=\Delta_+)$ \cite{Wi,GM,GK}.
In the IR limit conformal invariance is recovered and the
operator dual to
the scalar field $\phi$ has the
conformal dimension $\Delta_+$.
This does not cause any problems since the modes for both $\Delta_-$ and
$\Delta_+$ are normalizable due to our restriction (\ref{dimb}).

\subsection{Holographic Replica Method}

Now we are prepared to present the formulation of AdS/CFT for
disordered systems by employing the replica method\footnote{A slightly different replica method has
 been applied
to AdS/CFT in \cite{RT} to compute the entanglement entropy \cite{CC}.}.
If we consider a CFT with the random interaction (\ref{lagone}),
it is described by a deformation  of the product of
$n$ copies of the CFT,
$\text{CFT}_1\otimes \text{CFT}_2\otimes \ddd \otimes \text{CFT}_n$,
in the replica method.
Assuming the gaussian distribution of disorder (\ref{prob}),
the theory is deformed by double trace operators
as (\ref{cortwoo})
\be S_{int} =
-\f{f}{2}\int d^d x
\left(\sum_{i=1}^n {\cal O}_i(x) \right)^2, \label{intoo}
\ee
where we denote ${\cal O}_i$
as the operator ${\cal O}$ with its conformal dimension $\Delta$
in CFT$_i$. Compared with the standard double trace deformation (\ref{dti}), the sign of the coupling (\ref{intoo}) is opposite
(since $f$ in (\ref{prob}) must be positive),
a common feature of the replicated theory in disordered systems.
Thus, one may worry that this interaction may cause an instability in this theory.
If we wish,
we can add a conventional double trace deformation \eqref{dti}
to the original theory together with
the randomness (\ref{lagone}).
In the replica method, it means that we consider the product of
$n$ CFTs with the following generalized interaction
\be S_{int} =
-\f{f}{2}\int d^d x
\left(\sum_{i=1}^n {\cal O}_i(x) \right)^2
+\f{\lambda}{2}\int d^d x
\sum_{i=1}^n ( {\cal O}_i(x) )^2
. \label{into}
\ee
Later, however, we will see that the limit $\lambda =0$ of
two point functions is well-defined, and the two point
functions show more interesting scaling behaviors than the case
with $\lambda>0$.

In the dual AdS side, the spacetime is defined by $n$ copies of
an AdS space. They are disconnected in the bulk
and attached to the same boundary $R^{d}$ at $z=0$.
As seen above, the deformation by the double trace operator changes
the boundary condition for the dual field, therefore after the
deformation by \eqref{into} the fields in copies of an AdS space
interact with each other through the boundary conditions.
Interestingly, this setup with multiple AdS spaces has recently been
discussed in \cite{ACK,Ki} from a different motivation.
Originally we have $n$ gravitons dual to $n$ stress-energy tensors conserved independently.
After we put the interaction (\ref{into}),
only a combination of stress-energy tensors is conserved and hence
there exists only one dual massless graviton as argued in \cite{ACK,Ki}.
In the AdS gravity the other $n-1$ gravitons become massive due to the one-loop contribution,
while
in the dual CFT, the conformal dimension of other combinations of stress-energy tensors deviates from $\Delta=d$ to $d+\delta(n)$
due to the interaction (\ref{into}) \cite{Cardy,ACK,Ki}.
The standard bulk to boundary relation relates $\delta(n)$
to the mass of $n-1$ gravitons as $M^2_{g}=d\delta(n)$.

In the following, we calculate
two point functions at the tree level of AdS gravity in this setup.
We denote $\phi_i$ as the field
corresponding to ${\cal O}_i$, and
define $\ap_i(x)$ and $\beta_i(x)$ from the asymptotic behaviors
as in (\ref{expanda}).
Following the previous analysis, we obtain the deformed action as
\begin{align}
S[\beta,J]_n &=\int
d^d k \Biggl[\f{\sum_{i=1}^n\beta_i(k)\beta_i(-k)}{2G(k)}\ - \ \f{f}{2}
\left(\sum_{i=1}^n\beta_i(k)\right)\cdot\left(\sum_{i=1}^n\beta_i(-k)\right) \label{disac} \\
&\qquad \qquad \qquad
+ \f{\lambda}{2}\sum _{i=1}^n \beta_i (k)\beta _i (-k)
+\beta_1(-k)J_1(k)+\beta_2(-k)J_2(k)\Biggr]. \nonumber
\end{align}
Here we included the source terms only for $\beta_1$ and $\beta_2$ without losing generality,
by taking the symmetry into account. Notice that when $\lambda > 0$, this system (\ref{disac}) is stable
in the $n \to 0$ limit\footnote{\label{betarotation}%
Indeed, if we redefine $\hat \beta_0 =
\frac{1}{\sqrt{n}}\sum_i \beta_i$, $\hat \beta_i =
\beta_i - \frac{1}{n}\sum_j \beta_j$, then the interaction
terms are given by $\frac{1}{2}(\lambda - nf) \hat \beta_0^2
 + \frac{\lambda}{2} \sum_i \hat \beta_i^2$.
Therefore, any $\lambda>0$ would be enough to stabilize the
saddle points for small $n$.}.
Taking the limit of $\lambda\to 0$ in the end, we define
the theory with $\lambda=0$.

{}From the equations of motion $\delta S/\delta \beta_i=0$,
we find
\begin{align}
&\beta_1=\f{-fG(k)J_2(k)-(1+(\lambda +(1-n)f)G(k))J_1(k)}
{(1+\lambda G(k) )( 1 +(\lambda - nf) G(k) )}G(k),\no
&\beta_2=\f{-fG(k)J_1(k)-(1+(\lambda +(1-n)f)G(k))J_2(k)}
{(1+\lambda G(k) )( 1 +(\lambda - nf) G(k) )}G(k),\no
&\beta_3=\ddd=\beta_n= - \f{fG(k)^2(J_1(k)+J_2(k))}
{(1+\lambda G(k) )( 1 +(\lambda - nf) G(k) )} .
\end{align}
{}From these equations we can express the action $S$ as
\begin{align}
S[J]_n& =-\dfrac{1}{2}\int d^d k
\biggr( \dfrac{ G(k) \left( 1+(\lambda + f(1-n)) G(k) \right)
(J_1(k)J_1(-k)+J_2(k)J_2(-k))}{(1+\lambda G(k))(1+ (\lambda - nf )G(k))} \no
&\qquad \qquad \qquad \qquad
 +\dfrac{2fG(k)^2J_1(k)J_2(-k)}{(1+\lambda G(k))(1+ (\lambda - nf )G(k))} \biggr). \label{actjj}
\end{align}
While $n$ is a positive integer in our starting expression,
in (\ref{actjj}) we are free to regard $n$ as a continuous valuable,
which is a crucial assumption for the replica method.

Two point functions can be computed from \eqref{actjj} as
\begin{align}
& \la {\cal O}_1(k){\cal O}_1(-k)
 \lb_n={\frac { G(k) \left( 1+(\lambda + f(1-n)) G(k) \right) }{(1+\lambda G(k))(1+ (\lambda - nf )G(k))}}
, \label{cordis}\\
& \la {\cal O}_1(k){\cal O}_2(-k)
 \lb_n={\frac { f G(k)^2}{(1+\lambda G(k))(1+(\lambda - nf) G(k))}}
. \label{cordis2}
\end{align}
The subscript $n$ implies that the two point functions are
evaluated in the replicated theory with fixed $n$.
As discussed in section \ref{sec: Disordered Systems and Replica Method},
we take the limit $n \to 0$ and finally find in
the disordered system as
\be
\ov{\la {\cal O}(k){\cal O}(-k)\lb} = \la {\cal O}_1(k){\cal O}_1(-k)
 \lb= {\frac { \left( 1+(f+\lambda )G(k) \right) G(k)}{(1+\lambda G(k))^2}}.
\label{coronee}\ee
We can also compute
$\ov{\la {\cal O}(k)\lb \la {\cal O}(-k)\lb}$,
which involves, when replicated,  two distinct replicas,
\be
\ov{\la {\cal O}(k)\lb \la {\cal O}(-k)\lb}
=\la {\cal{O}}_1(k){\cal{O}}_2(-k)\lb ={\frac { f G(k)^2}{(1+\lambda G(k))^2}}.
\label{cortwo}
\ee
Unlike
$\ov{\la {\cal O}(k){\cal O}(-k)\lb}$,
$\ov{\la {\cal O}(k)\lb \la {\cal O}(-k)\lb}$ is made non-zero solely
because of disorder,
and hence our analysis indeed describes a disordered phase.
Correlation functions of this type have been used as an order parameter
in spin glass theories
\cite{Binder_Young_RMP}.

When $\lambda>0$,
we can observe
from the two point function \eqref{coronee}
that the operator ${\cal O}$ with conformal dimension $\Delta$
always flows into the one with
$\Delta_+ = d - \Delta$ in the IR limit $k\to 0$.
Therefore, the IR limit of the disordered system looks
similar to the theory deformed by
a single double trace operator \eqref{dti}%
\footnote{Fixed points for more generic cases with $n=2$ are
analyzed in \cite{Ki} and their result is consistent with ours.}.
In the appendix \ref{Order/Disorder Phase Transition},
we suggest that the random spin system, such as
the random bond Ising model \cite{random_bond_Ising}, is
analogous to this case with $\lambda=f$, where the
order/disorder phase transition occurs.
In the appendix \ref{apb}, we analyze
the effective central charge $c_{eff}$ \cite{Cardy,CL}
of this disordered CFT and calculate the difference
between
the effective central charge
at the UV (trivial) fixed point ($c^{UV}_{eff}$)
and at the IR fixed point ($c^{IR}_{eff}$) .
We find that the inequality $c^{IR}_{eff}-c^{UV}_{eff}\leq 0$ always
holds for any values of $\lambda>0$.
This may support an analogue of $c$-theorem in the disordered system,
though there has been no proof of the $c$-theorem for $c_{eff}$ from the CFT side.

At the special point $\lambda =0$, we can obtain
a markedly different
result. In the IR limit $k\to 0$, the operator ${\cal O}$
with dimension $\Delta$ becomes
an operator with dimension $2\Delta-d/2$, as seen
from the behavior
\be
\ov{\la {\cal O}(k){\cal O}(-k)\lb} \sim k^{4\Delta-2d}.
\ee
The constraint (\ref{dimb}) leads to $d/2-2\leq
2\Delta-d/2\leq d/2$, and hence
the lower bound of $2\Delta-d/2$
 violates the unitarity bound.
However, this may be fine since the disordered system is not a closed unitary system.
The analysis of the effective central charge $c_{eff}$ at $\lambda=0$
seems to need a special care
as discussed in the appendix \ref{apb}
and we leave it as a future problem.
Observe that the IR limit is not fully gapped, but rather
the theory flows into another critical field theory.
As it becomes clear from the RG analysis in the next section, the random fixed
point corresponds to the infinite randomness limit ($f\to \infty$).
This behavior could be comparable with
random vector potential models of
Anderson localization
\cite{random_vector}
or
random quantum spin chains
\cite{random_singlet}.

In principle, we can extend our computations to higher point functions by including interaction
terms in (\ref{actjj}). In the case of a cubic coupling, we can add to (\ref{disac}) a cubic term like
\be
S_{cubic}=\int d^d k_1 d^d k_2 \,
C(k_1,k_2) \, \beta(k_1) \, \beta(k_2) \, \beta(-k_1-k_2), \label{cubic}
\ee
where $C(k_1,k_2)$ is related to the three point function in momentum space for $f=0$ via
the standard rule \cite{GKP,WAdS}.

It is also possible to add
more general multi-trace
interactions $\int d^d x\, W ({\cal{O}}(x) )$ in the boundary CFT.
Correspondingly,
in the holographic description, we need to add
the potential term $\int d^d x \, W(\beta(x))$ to (\ref{disac}), which leads to the generalized action
\be
S= K\int d^d x d^d y \f{\sum_{i=1}^n\beta_i(x)\beta_i(y)}{|x-y|^{2(d-\Delta)}}+\int d^d x \left[
-\f{f}{2}
\left(\sum_{i=1}^n\beta_i(x)\right)^2+\sum_{i=1}^nW(\beta_i(x))\right],\label{generalp}
\ee
where $K$ is a numerical constant.

Similarly, we can consider more general probability
distributions for disorder other than the gaussian white-noise.
After the theory is replicated, it
leads to a deformation\\
\begin{align}
(-1)\sum^{\infty}_{p=1} \frac{(-1)^p}{p!}
\int d^d x_1
\cdots
\int d^d x_p
\overline{ g(x_1)\cdots g(x_p)}^c
\sum_{a_1,\ldots,a_p}
\mathcal{O}_{a_1}(x_1)
\cdots
\mathcal{O}_{a_p}(x_p) ,
\end{align}
where
$\overline{ g(x_1)\cdots g(x_p)}^c$
is the $p$-th cumulant of the probability distribution $P[g(x)]$.

\subsection{Examples of Disorder Systems in AdS/CFT}

\subsubsection{Random ${\cal N}=4$ Super Yang-Mills}

As the first example,
we can consider the disordered system of
${\cal N}=4$ super Yang-Mills theory in four dimensions by a random deformation (\ref{lagone}).
We can take ${\cal O}$ to be a 1/2 BPS operator with $\Delta=2$ made of two transverse scalars, i.e.
${\cal O}_{ab}=\mbox{Tr}[\Phi^a\Phi^b]-\f{1}{6}\delta^{ab}\mbox{Tr}[\Phi^c\Phi^c]$. In this case,
 the condition (\ref{relevant}) is saturated and therefore the RG flow
is logarithmic.

\subsubsection{Random $O(N)$ Magnet}\label{ronm}

Klebanov and Polyakov \cite{KP} conjectured that the massless fields in $AdS_4$
with even spins describe the singlet sector
of the three-dimensional
critical $O(N)$ vector model in the large $N$ limit.
By using the vector field $\vec{\phi}$ with $N$ components,
the action of the $O(N)$ vector model is
\be
S=\f{1}{2}\int
d^3x \left[\de\vec{\phi}\cdot \de\vec{\phi}
+\f{\lambda}{2N}(\vec{\phi}\cdot \vec{\phi})^2\right].\label{ona}
\ee
There are two critical points; One is at $\lambda=0$, i.e.,
 the free field theory. The other is at the end point of the RG flow induced by the second term in (\ref{ona}), which is interpreted
as a double trace deformation by setting ${\cal O}=\vec{\phi}\cdot \vec{\phi}$
in section \ref{Double Trace Deformation in AdS/CFT}.
Notice that the dimension of ${\cal O}^2=(\vec{\phi}\cdot \vec{\phi})^2$ is $2\Delta=2$ and thus
it is relevant. In the IR fixed point,
the conformal dimension of ${\cal O}$
changes into $d-\Delta=2$ and then ${\cal O}^2$ becomes irrelevant.

Now, starting from the $O(N)$ vector model,
we introduce the randomness via the interaction
$\int d^3x g(x) \, \vec{\phi}(x)\cdot \vec{\phi}(x)$.
In the replica method, this disordered system is described by the $n\to 0$ limit of the following
system
\be
S=\f{1}{2}\int d^3 x
\left[\sum_{i=1}^n\de\vec{\phi}_i\cdot \de\vec{\phi}_i-\f{f}{N}\sum_{i,j=1}^n
(\vec{\phi}_i\cdot \vec{\phi}_i)(\vec{\phi}_j\cdot \vec{\phi}_j)
+\f{\lambda}{2N}\sum_{i=1}^n
(\vec{\phi}_i\cdot \vec{\phi}_i)^2\right].\label{onmodel}
\ee
Its holographic description is precisely given by the model (\ref{disac})
analyzed in the previous subsection.
We then conclude that whenever $\lambda \neq 0$, 
disorder is innocuous in the IR limit. 
In particular, disorder is an irrelevant perturbation
at the non-trivial fixed point of (\ref{ona}),
which agrees with the well-known fact 
for the 3D $O(N)$ magnets with $N\ge 2$. 
(See, for example, 
\cite{Pelissetto_Vicari00,Pelissetto_Vicari_review}.)

\subsection{Comments on Replica Symmetry Breaking}

The replica symmetry is the symmetry
under a permutation of fields
among different replicas,
such as $\beta_i \to \beta_j \ (i\neq j)$ in our setup.
The previous analysis (\ref{disac}) clearly preserves the replica symmetry
as all of  $\beta_i$ are vanishing. The breaking of the replica symmetry
typically occurs when there are many vacua in the replica theory.
If this happens, the analysis becomes more
non-trivial because
the definition of order parameter (or mean field) of randomness gets complicated.
A famous such
example is the problem of the spin-glasses
(see e.g. \cite{Binder_Young_RMP}).

Indeed, we can find examples where the replica symmetry is spontaneously broken
in our holographic setup as follows. If we are interested in the IR
limit, we can drop off the first term in (\ref{disac}) and it is
straightforward to study the vacuum structure of this system from the potential terms. By choosing $W(\beta)$ appropriately\footnote{
For example, we can realize this situation
if we assume $\f{dW(\beta)}{d\beta}$ is a oscillating function so that its average is increasing and
$\f{dW(\beta)}{d\beta}=0$ has only a single solution $\beta=0$.},
we can realize situations where the potential $\sum_{i}W(\beta_i)$
has only the trivial vacua\footnote{Here we include
metastable vacua into our definition of vacua.}%
, i.e., $\beta_i=0$,
while the potential with the term proportional to
$f$ has multiple vacua with $\beta_i \neq \beta_j$.

\section{Field Theory Analysis in the Planar Limit}
\label{Field Theory Analysis in the Planar Limit}

In this section, we show
that the two point functions \eqref{coronee} and \eqref{cortwo}
obtained from
our holographic method
can be reproduced from field theoretic calculations
in the planar limit. This confirms the
validity of our formulation of the holographic replica method. Before considering
the replicated case,
we reproduce
from the field theory calculations the
two point function
\eqref{twopp} in the case of the deformation of \eqref{dti}
\begin{align}
S_{int}= \lambda \int d^d x \, \Phi_{\rm pert} (x) , \qquad
  \Phi_{\rm pert} (x) = \frac12 ( {\cal O} (x) )^2 .
\end{align}
In order to consider the renormalization of $\lambda$ and $\cal O$,
we need to compute beta function $\beta_\lambda$ and anomalous dimension
$\gamma_{\cal O}$. In the large $N$ limit, non-trivial contributions
come only from the two point function
$\langle {\cal O} (x) {\cal O}(0) \rangle
 = v/ |x|^{2 \Delta}$ with
$v= \pi^{-d/2} \frac{(2\Delta - d)\Gamma(\Delta)}{\Gamma(\Delta-d/2)}$,
and those from higher point functions are suppressed
(see, e.g., \cite{Wi}).
Therefore, divergent terms arise only from the following
operator product expansions (OPEs) as
\begin{align}
  \Phi_{\rm pert} (x) \Phi_{\rm pert} (0) \sim
   \frac{2 v}{|x|^{2 \Delta}} \Phi_{\rm pert} (0) ,
   \qquad
    \Phi_{\rm pert}(x) {\cal O} (0)
   \sim \frac{v}{|x|^{2 \Delta}}
  {\cal O} (0) .
\end{align}
{}From these OPEs we can obtain the beta functions following a standard
analysis of quantum field theories\footnote{For an extensive analysis of RG flows in the $n=2$ case,
refer to the third paper in \cite{Ki}.}
\begin{align}
 \frac{d}{d \ln |k|} \tilde \lambda (k) = \beta_{\tilde \lambda} =
   ( 2 \Delta - d) \tilde \lambda (k)
  + (\tilde \lambda (k))^2 , \qquad
  \gamma_{\cal O} = \Delta + \frac12 \tilde \lambda (k),
  \label{beta}
\end{align}
where we redefine
$\tilde \lambda = \frac{2^{d-2\Delta}(2\Delta - d) \Gamma(d/2 - \Delta)}
{\Gamma(\Delta - d/2)} \lambda$.
Notice that even though
we included only the leading order corrections
to
$\lambda$,
the result is exact in the large $N$ limit.
The beta function \eqref{beta} leads to
\begin{align}
 \tilde \lambda (k) = \frac{ (d - 2 \Delta ) \tilde \lambda_0}
 { |k|^{ d - 2 \Delta } + \tilde \lambda_0 } .
\end{align}
Since the RG flow equation is given by
\begin{align}
 \left [ \frac{d}{d \ln |k|} - \beta_{\lambda} \frac{d}{d \lambda}
  + d - 2 \gamma_{\cal O} \right ]
 \langle {\cal O} (k) {\cal O} ( - k) \rangle = 0 ,
 \label{rgflow}
\end{align}
the two point function is
\begin{align}
\langle {\cal O} (k) {\cal O} ( - k) \rangle
 = {\cal C}
 \exp \left( - \int^{\ln |k|} d \ln |k '|
 (d - 2 \gamma_{\cal O} (k ' ))
  \right) = \frac{G(k)}
 {1 + \lambda_0 G(k) }.
\end{align}
The coefficient ${\cal C}$ is fixed such that
$\langle {\cal O} (k) {\cal O} ( - k) \rangle = G(k)$ for
$\lambda_0=0$.

Next let us introduce furthermore disorder \eqref{lagone} with
the gaussian distribution \eqref{prob}.
In the replica method the disordered system can be represented by
introducing $n$ copies of the CFT with ${\cal O}_i
(x)$ $(i=1,2,\cdots ,n)$
whose OPEs
are ${\cal O}_i (x) {\cal O}_j
(0) \sim
 \delta_{i,j} v / |x|^{2 \Delta}$.
Moreover we deform the $n$ copies of the CFT by \eqref{into}
 \begin{align}
  & S_{int}
= - f \int d^d x \, \Phi_\text{pert} (x) +
  \lambda \int d^d x \, \Psi_\text{pert} (x)
 , \\
  &\Phi_\text{pert} (x) =
 \frac12 \left( \sum_{i=1}^n {\cal O}_i (x) \right)^2 , \qquad
  \Psi_\text{pert} (x) = \frac12 \sum_{i=1}^n ( {\cal O}_i (x) )^2 .
\end{align}
As in the previous analysis, we now
reproduce the two point functions \eqref{coronee} and \eqref{cortwo}.
From the OPE coefficients, we read off the beta functions as%
\footnote{
When specialized to the case of the random $O(N)$ magnet
discussed in section \ref{ronm},
the beta functions
are consistent with the known results to the leading order in $N$. 
See, for example, 
\cite{Pelissetto_Vicari00,Pelissetto_Vicari_review}.
}
\begin{align}
 &\frac{d}{d \ln |k|} \tilde f (k) = \beta_{\tilde f}
 = (2 \Delta - d) \tilde f (k) - n  (\tilde f (k) )^2
 + 2 \tilde f (k) \tilde \lambda (k),
\nonumber 
\\
 &\frac{d}{d \ln |k|} \tilde \lambda (k) = \beta_{\tilde \lambda}
 = (2 \Delta -d ) \tilde \lambda (k) + ( \tilde \lambda (k) )^2.
\label{beta fn}
\end{align}
We find, therefore,
\begin{align}
 \tilde f(k) - \frac{ \tilde \lambda (k)}{n}
 = \frac{(d - 2 \Delta ) ( \tilde f_0 - \tilde \lambda_0 / n )}
 { |k|^{d - 2 \Delta} + \tilde \lambda _0 - n \tilde f_0} , \qquad
\tilde \lambda (k) = \frac{(d - 2 \Delta) \tilde \lambda_0 }
 { |k|^{d - 2 \Delta} + \tilde \lambda_0 } .
\end{align}
For computing the anomalous dimensions
it is convenient to rotate the operators as
\begin{align}
 \hat {\cal O}_0 (x ) = \frac{1}{\sqrt{n}}
 \sum_{i=1}^n {\cal O}_i (x) ,
 \qquad
  \hat {\cal O}_j (x ) =
  {\cal O}_j (x) - \frac{1}{n} \sum_{i=1}^n {\cal O}_i (x) ,
 \label{es}
\end{align}
with $j=1,\cdots ,n$, according to the irreducible
 representation of symmetric group
${\cal S}_n$. Here the number of independent operators does not change
since $\sum_{j=1}^n \hat {\cal O}_j = 0$.
In this normalization we have $\hat {\cal O}_0 (x) \hat
{\cal O}_0 (0) \sim v / |x|^{2 \Delta}$
and $\hat {\cal O}_i (x) \hat
{\cal O}_j (0) \sim (\delta_{i,j} - \frac{1}{n}) v /
 |x|^{2 \Delta}$.
The anomalous dimensions of
new operators are obtained just like before as
\begin{align}
 \gamma_{\hat {\cal O}_0} = \Delta + \frac{1}{2} \tilde \lambda
 - \frac{n}{2} \tilde f , \qquad
 \gamma_{\hat {\cal O}_j} = \Delta + \frac{1}{2} \tilde \lambda ,
\end{align}
and solving the RG flow equation we have
\begin{align}
 \langle \hat {\cal O}_0 (k) \hat {\cal O}_0 ( -k) \rangle_n
 = \frac{G(k)}
 {1 + ( \lambda_0 - n f_0) G(k)} , \qquad
 \langle \hat {\cal O}_i (k) \hat {\cal O}_j ( -k) \rangle_n
 = \left(\delta_{i,j} - \frac{1}{n} \right) \frac{G(k)}{1 + \lambda_0 G(k)} .
\end{align}
Rotating the operators back again, now we can reproduce the
previous results of the two point functions
\eqref{cordis} and \eqref{cordis2}, therefore after taking
$n \to 0$ limit we have
\eqref{coronee} and \eqref{cortwo}.

The method with the RG flow equation might be a standard way to compute
correlation functions, but there is another way
in a field theory viewpoint via a Hubbard-Stratonovich transformation first considered by
Gubser and Klebanov
\cite{GK}. The partition function we consider
can be rewritten as
\begin{align}
 Z^n_f [J] &=
 \left(\det - \frac{1}{f}\right)^{\frac{1}{2}}
  \left(\det \frac{1}{\lambda}\right)^ {\frac{n}{2}}
\int D g(x) \prod_{i=1}^n D \sigma_i(x)
   \times \\
  & \qquad \times \left\langle e^{ \int d^d x [-\frac{1}{2f} g (x) ^2
 + \frac{1}{2\lambda} \sum_{i=1}^n \sigma_i^2
 + \sum_{i=1}^n ( g (x) + \sigma_i (x) + J_i (x) ) {\cal O}_i(x) ]} \right\rangle _0 ,\nonumber
\end{align}
where the subscript $0$ suggests that the correlation function is
computed without perturbation, i.e. with $f=\lambda = 0$.
Integration over $\sigma_i$ reproduces the double trace deformation
$\frac{\lambda}{2} \int d^d x \, \sum_i {\cal O}_i^2$.
Notice that the
non-trivial contribution arises only from the two point function $\langle
{\cal O}_i(x) {\cal O}_j(0) \rangle = \delta_{i,j} G(x)
 \, (\, \equiv \delta_{i,j} v/|x|^{2 \Delta})$ in the
large $N$ limit. {}From this observation we find
\begin{align}
 Z^n_f [J] &=
  \left(\det - \frac{1}{f}\right)^{\frac{1}{2}}
  \left(\det \frac{1}{\lambda}\right)^ {\frac{n}{2}}
 \int D g(x) \prod_{i=1}^n D \sigma_i(x)
   \times \\
  & \qquad \times
 e^{\int d^d x [-\frac{1}{2f} g (x) ^2 + \frac{1}{2\lambda}
 \sum_{i=1}^n \sigma^2 _i(x)
 + \frac{1}{2} \sum_{i=1}^n
 ( g (x) + \sigma_i (x) + J_i (x) ) \hat G ( g (x) + \sigma_i(x) + J_i(x) ) ]} ,\nonumber
\end{align}
where $\hat G g (x) = \int d^d y \, G(x - y) g(y)$. Since this
expression is gaussian with respect to $g$ and $\sigma_i$,
we can easily integrate them out. The result is%
\footnote{The prefactor corresponds to the $O(N^0)$ corrections
to the partition function, and it can be used to compute the shift
of central charge along the RG flow.
See, for example, \cite{GK} and appendix \ref{apb}.}
\begin{align}
 Z^n_f [J] &= (1 + \lambda \hat G)^{-\frac{n-1}{2}}
  (1 + (\lambda - nf ) \hat G)^{-\frac{1}{2}}
   \times \\
  & \qquad \times
  e^{\frac{1}{2} \int d^d x [ \sum_{i=1}^n
 J_i (x) \hat Q J_i (x) + (\sum_{i=1}^n J_i (x) )
   \hat Q ' (\sum_{i=1}^n J_i (x) ) ] } , \nonumber
\end{align}
with
\begin{align}
  \hat Q = \frac{\hat G}{1 + \lambda \hat G} , \qquad
  \hat Q ' = \frac{ f \hat Q ^2}{1 - n f \hat Q} .
\end{align}
Taking derivatives with respect to $J_i$ twice, we obtain
the two point functions \eqref{cordis} and \eqref{cordis2}
and again reproduce \eqref{coronee} and \eqref{cortwo} in
the $n \to 0$ limit.

\section{Conclusions and Discussions}
\label{conclusion}

In this paper, we have studied the quenched disordered systems
with arbitrary strength of disorder
by applying the AdS/CFT correspondence. We formulate the holographic
replica method by employing the setup of the double trace deformation in AdS/CFT.
We have calculated the two point functions in our disordered system in the planar limit from both the AdS and CFT sides and got
 the same result. We found that
the scaling of the two point functions evolves non-trivially
under the RG flow. Especially, if we fine tune the parameter to $\lambda=0$, then the
two point functions in the IR limit show remarkably new behavior. In the generic case
$\lambda>0$, the IR limit is essentially the same as the fixed point obtained by the standard double trace
deformation.
As in appendix \ref{apb} it is possible to analyze the effective
central charge $c_{eff}$ using the AdS/CFT correspondence.
There we observe that it decreases
under the RG flow between two fixed points when $\lambda>0$. A more thorough analysis of this would
be very intriguing as no proof of the $c$-theorem about $c_{eff}$ has been known so far.

It is important to remember that
the subtle limit $n\to 0$ of the replica
method does not cause any problems in our examples. We also pointed out that the replica symmetry
may be broken if we consider particular deformations of the CFT. It may also be intriguing to
note that the limit $n\to 0$ offers us a formal way to construct AdS duals to non-unitary CFTs
with the central charge $c=0$.

Even though our holographic formalism of the replica method covers quantum disordered systems, our
explicit computations are performed only
for classical disordered systems. Thus
here we would like to mention the application of our holographic replica method to
quantum disordered systems. Since the random coupling $g(x)$ does not depend on the time $t$,
the frequency $\omega$ of the field $\beta_i$ in (\ref{disac}) should be vanishing.
Thus the results \eqref{coronee} and \eqref{cortwo} remain the same only when $\omega=0$; Otherwise
we will get the same result as the one in the pure system, i.e., with
$f=0$. This triviality
is because we are taking the planar limit in the presence of the multi-trace interactions. In order to obtain
$\omega$-dependent results, we need to go beyond the tree level analysis in the gravity theory.
A similar situation occurs when we are considering two point functions of other operators, such as the currents $J^\mu(t,x)$. At the tree level, we do not find any $f$-dependence in both classical and
quantum disorder systems. These problems of one loop analysis clearly deserve
future studies.

In addition to the replica method, we know another method which
also enables us to study disordered systems, called the supersymmetric method, as mentioned before. In order to
deal with the randomness in a theory with scalar fields and Dirac fermions we add their
superpartners (or ghosts). In this procedure, the bosonic global symmetry, such as $O(N)$, becomes
its supergroup extension, such as $OSp(N|N)$. It may be interesting to
find a similar method for Yang-Mills gauge
theories. Naively, one may think that we can replace the bosonic gauge group $U(N)$ with the
supergroup $U(N|M)$ (see \cite{OT} for gauge theories with supergroup gauge symmetries).
However, we can easily see that this leads to what we precisely want only
at the vanishing coupling $g_{YM}=0$. Therefore we need a modification or
another idea.

\vskip5mm

\noindent {\bf Acknowledgments}

We would like to thank T. Azeyanagi, M. Fukuma, S. Hartnoll, C. Herzog, S. Iso,
J. McGreevy and T. Nishioka for useful conversations. We are also grateful to E. Kiritsis
for a useful comment.
The work of YH is supported by JSPS Research Fellowship.
SR thanks Center for Condensed Matter Theory
at University of California, Berkeley for its support.
The work of TT is supported by JSPS Grant-in-Aid for Scientific Research
No.18840027 and by JSPS Grant-in-Aid for
Creative Scientific Research No. 19GS0219.

\appendix

\section{Order/Disorder Phase Transition}
\label{Order/Disorder Phase Transition}

In condensed matter systems,
randomness often competes with ordering tendencies.
For example, for a spin system (e.g. the Ising model) with critical temperature $T_c$,
the system is in an ordered phase (e.g. ferromagnetic phase) for $T<T_c$.
Even if we introduce a small amount of disorder
the system will still be in the ordered phase.
However, if the randomness becomes strong
enough, a phase transition will occur and
the system will evolve into a disordered phase. This is a
standard story, for example, in the ferromagnets
in the presence of random magnetic field (see e.g. \cite{CardyBook}).

Such competition can be described in our holographic approach.
In the spin system examples, we regard ${\cal O}$ as the spin operator $\sigma$.
Then an important point is that we omit the $i=j$ terms in the interaction
terms $-\f{f}{2}\sum_{i,j}{\cal O}_i{\cal O}_j$ (see \eqref{into})
since the operator $:{\cal O}^2:$ does not exist in spin systems like
the Ising model%
\footnote{For example, if we consider the Ising model,
the OPE of the spin operators is $\sigma\cdot\sigma= I+{\cal{E}}$.
This OPE produces the energy density operator ${\cal E}$, but
this just shifts the critical temperature. In contrast,
if we consider the large $N$ $O(N)$ vector
model, then the operator $:{\cal O}^2:$ exists and we have to keep it as in (\ref{disac}).
Even in the latter case, we can add the interaction term as in
\eqref{into} with $\lambda = f$ and realize a system without
the $i = j$ terms.}.
Another point is that when $T<T_c$ and $f=0$, the system is in a ordered phase.
As an example we express this with the spontaneous breaking of the
$Z_2$ symmetry $\beta\to -\beta$ by adding the standard
wine bottle potential $W(\beta)=-m^2\beta^2+\lambda \beta^4$. Notice that $m^2>0$ when $T<T_c$, while
$m^2=0$ at $T=T_c$. In summary, we reach the following action of $\beta$
\be
S[\beta]_n =\int
d^d k \Biggl[\f{\sum_{i=1}^n\beta_i^2}{2G}\ - \ \f{f}{2}
\sum_{i\neq j}^n\beta_i\beta_j+ \sum_{i}(-m^2\beta_i^2+\lambda \beta_i^4)\Biggr].\label{acx}
\ee
Assuming that the vacuum preserves the replica symmetry,
i.e., $\beta_1=\beta_2=\ddd=\beta_n\equiv\beta$,
we can rewrite (\ref{acx}) as follows
\be
S[\beta]_n =n\int d^d k\Biggl[\f{1}{2G}\beta^2-\f{f}{2}(n-1)\beta^2-m^2\beta^2+\lambda \beta^4\Biggr].
\ee
In the $n\to 0$ limit, the term due to the randomness behaves like $\f{f}{2}\beta^2$ and it competes with
the spontaneous breaking of the $Z_2$ symmetry. Therefore we can realize the disorder/order
phase transition when we increase the randomness parameter $f$ in our holographic description.

\section{Effective Central Charges in Disordered Systems}\label{apb}

We can measure the degrees of freedom of a given CFT in even dimensions by
calculating the central charge.
Let us suppose that the replica theory has a non-trivial fixed point.
Then we can define the central charge of this new CFT as
\begin{align}
 c(n) = n c_0 + \Delta c (n),
\end{align}
for any fixed $n$.
Here $c_0$ is the central charge of the original CFT without
any deformations. For example, in the ${\cal N}=4$ $U(N)$ gauge theory case it is given by $c_0=N^2 / 4$.
The term $\Delta c(n)$ is due to the deformations and comes from one-loop order corrections $O(N^0)$.
In the random system, we can still define so called the effective central charge $c_{eff}$ (see e.g.\cite{Cardy,CL})
\be
c_{eff}\equiv \left. \f{dc(n)}{dn} \right|_{n=0}. \label{eccf}
\ee
Physically, this central charge is equal to the coefficient of the OPE of energy stress tensors at the random fixed point \cite{Cardy}
\be
 \overline{\la T_{\mu\nu}T_{\mu'\nu'}\lb} \propto c_{eff} .
\ee
In a usual unitary CFT, there is a $c$-theorem, which states that
the central charge is decreasing under the RG flow.
However, in disordered systems,
no $c$-theorem is known for $c_{eff}$
even if the replicated theory with fixed $n$
is unitary, because finally we need to
take the formal $n\to 0$ procedure.

In our setup of AdS/CFT, the order $O(N^0)$ correction to
the central charge is proportional to the one from the
one-loop effective potential
$V_\text{1-loop}=-\f{1}{2}\mbox{Tr}_\text{AdS}(-\square+m^2)$
of the scalar field.
Since our formulation of holographic replica model is closely
related to the case with the double trace deformation \eqref{dti},
let us first review
the case analyzed in \cite{GM}.
In the case with the double trace deformation (\ref{dti}),
the $\lambda$-dependence of $V_\text{double}$ comes from the boundary condition for the field $ \phi(x,z)$ imposed at the boundary $z = 0$.
For $\alpha (x)$ and $\beta (x)$ in (\ref{expanda}), we assign
\be
\ap(x)=- \lambda \beta(x),
\ee
which can be obtained from the relations (\ref{eomj}) and (\ref{abg}).
The one-loop vacuum
energy was explicitly calculated in \cite{GM} as
\be
V_\text{double}(\lambda)=-\f{1}{2^{d-2}\pi^{d/2}\Gamma(d/2)R^{d+1}}\int^\nu_0 d\ti{\nu}
\f{\ti{\nu}}{\Gamma(\ti{\nu})\Gamma(1-\ti{\nu})}\int^\infty_0 dp\f{p^{d-1}
\ti{\lambda}}{p^{2\ti{\nu}}+\ti{\lambda}}K_{\ti{\nu}}(p)^2,\label{vdb}
\ee
where $\nu=\f{d}{2}-\Delta$ and $\ti{\lambda}=2^{2\nu}\f{\Gamma(1+\nu)}{\Gamma(1-\nu)} \lambda z^{2\nu}$.
Here $z$ is the radial coordinate and $R$ is the radius of $AdS_{d+1}$.
 In the above expression, we subtracted the divergent piece which does not depend
on $f$, and which would actually
be canceled by other contributions due to the supersymmetry.
{}From this result we can show that
$V_{\mathrm{double}}(\lambda=\infty)<0$, and this is
consistent with the $c$-theorem as it is proportional to
$c_{IR}-c_{UV}$.

Now we compute the one-loop vacuum energy for the product of $n$ copies
of the CFT
with the deformation \eqref{into}.
In this case the boundary conditions are modified as
\be
\ap_j=f\sum_{i=1}^n \beta_i -\lambda \beta_j \label{condb}
\ee
for $j=1,2,\ddd,n$, or performing the redefinitions as
in footnote \ref{betarotation},
\be
\hat \ap_0 = - (\lambda - nf ) \hat \beta_0 ,\quad
\hat \ap_j=-\lambda \hat \beta_j .
\ee
Here we should notice that
$\sum_j \hat \alpha_j = \sum_j \hat \beta_j = 0$ by construction.
Therefore, by comparing with the previous result, we find
\be
V_\text{replica}(f,\lambda,n)=V_\text{double}\left(\lambda-nf \right)+(n-1)V_\text{double}(\lambda). \label{potvr}
\ee
Notice that the identity $V_\text{replica}(f,\lambda , n= 0)=0$
holds  as expected.
What we are interested in is the shift of the effective central charge $c_{eff}$ (\ref{eccf})
under the RG flow%
\footnote{Strictly speaking, the notion of central charge can be
used only at the conformal points. Therefore we should regard
\eqref{dcfn} as a definition of
an analogue of Zamolodchikov's $c$-function.};
\be
\Delta c_{eff} =  \left. \f{d\Delta c(n)}{dn} \right|_{n=0}=A\left(V_\text{double}(\lambda)
-f\cdot V'_\text{double}(\lambda)\right),
\label{dcfn}
\ee
where $A$ is a positive constant\footnote{It is explicitly given by
$A=c_0\f{8\pi G^{(5)}_NR^2}{2d}$.
}.

We would like to calculate the difference between $c_{eff}$  at the UV fixed point
$\lambda=f=0$ and at the IR fixed point.
If we assume $\lambda>0$, then $c_{eff}$ of the IR fixed point can be obtained
by setting $\lambda=\infty$ (recall the analysis of
the RG flow
in section \ref{Field Theory Analysis in the Planar Limit}).
Then we immediately find
\be
c^{IR}_{eff}-c^{UV}_{eff}=A\cdot V_\text{double}(\infty)<0. \label{cthr}
\ee
This seems to support the
$c$-theorem like property for the effective central charge of our disordered CFT.
However, we would like to stress again
that the $c$-theorem for $c_{eff}$ has not been proven at present and
a counter example may be found in more generic cases.
This value of $c^{IR}_{eff} - c^{UV}_{eff}$ (\ref{cthr}) is
actually the same as that of the standard double trace deformation \cite{GM,GK}
even in the presence of the random perturbation.

Moreover, if we restrict ourselves to the case with
$0<f<\lambda$, then we can show that $\Delta c_{eff}$
is always non-positive for any values of $f$ and $\lambda$ from the expression
(\ref{vdb}). Therefore we conclude that $\Delta c_{eff}$ is a monotonically decreasing function of
$f$ and $\lambda$ in this case.

In the remaining special case $\lambda=0$,
we would get naively
the opposite result $c^{IR}_{eff}-c^{UV}_{eff}>0$.
However, this analysis
is not reliable because $V_\text{double}\left(-nf \right)$ in (\ref{potvr}) is divergent for any values
of $n>0$. This subtlety arises since at $\lambda = 0$ the system of
(\ref{disac}) becomes unstable as mentioned before. We may instead deform the theory by, for example, a quartic term $\beta^4$ so that it become
stabilized. We would like to leave the analysis of $c_{eff}$ at $\lambda=0$ as a future problem.

\end{document}